\documentclass{aa}  
\usepackage[table]{xcolor}
\usepackage{multirow}
\usepackage{multicol}
\usepackage{colortbl}
\usepackage{graphicx}
\usepackage{txfonts}
\definecolor{CyanBG}{rgb}{0.71,0.85,0.93}
\usepackage{graphicx}
\usepackage{txfonts}
\usepackage{hyperref}

\begin{document}

   \title{The possible origin of three Apollo asteroids}
   \subtitle{3200 Phaethon, 2005UD, and 1999YC}

    \author{N. Kne\v zevi\' c \inst{1} \fnmsep \inst{2}
            \and  
            N. Todorovi\' c \inst{1}}
          
    \institute{
               Astronomical Observatory, Volgina 7, 11060 Belgrade, Serbia\\
               \email{nknezevic@aob.rs, ntodorovic@aob.rs}
               \and
               Department of Astronomy, Faculty of Mathematics, University of Belgrade, Studentski trg
               16, 11000 Belgrade, Serbia
              }

   \date{Received May 9, 2024; accepted June 7, 2024}

   \abstract
    {} 
   {We study the possible dynamical background of three Apollo asteroids: 3200 Phaethon, 2005 UD, and 1999 YC. The source regions under consideration are the asteroid families (2) Pallas, in the outer belt, and two inner-belt families (329) Svea and (142) Polana. We also aim to explain some of the contradictions in the literature in regards to the origin of Phaethon.}
   {Our methodology relies on the precise dynamical mapping of several mean motion resonances (MMRs), which are considered the main transport channels. This approach allows the clear detection of chaotic structures in an MMR and efficent selection of test asteroids for diffusion. We tracked the orbital evolution of the selected particles over 5 million years and registered all their eventual entries into the orbital neighborhood of the asteroids 3200 Phaethon, 2005 UD and 1999 YC. We performed massive calculations for different orbital and integration parameters using Orbit9 and Rebound software packages.}
   {We observed possible connections between three targeted Apollo asteroids and asteroid families we considered as their sources. The (2) Pallas family has the highest chance of being the origin of targeted asteroids, and (142) Polana has the lowest. The amount of transported material largely depends on the integrator, the integration step, and even the choice of the initial epoch, though to a lesser extent. There is a systematic discrepancy between the results obtained with Orbit9 and Rebound regarding the efficiency of the transport, but they show good agreement over delivery times and dynamical maps. A non-negligible number of objects approached all three target asteroids, which could indicate that the breakup of the precursor body occurred during its dynamical evolution.}

   {}


   \maketitle

\section{Introduction}

Phaethon (3200) is one of the most prominent asteroids. It is located in the Apollo region on a highly eccentric ($e\sim 0.89$) and inclined ($i \sim 22.28^\circ$) orbit, and is classified as a B-type asteroid \citep{Licandro2007}, with an estimated size of between 4 and 7 km \citep{Veeder1984, Green1985, Tedesco2002, Hanus2016, Taylor2019, Herald2020,  Kiselev2022}. Phaethon was the first discovered asteroid associated with a meteor stream: the Geminides \citep{GreenKowal1983, Whipple1983, FoxWillHug1983, Hughes1983, BabadzhanovObrubov1986, WilliamsWu1993}.

There have been countless studies of Phaethon, many of which addressed questions such as whether it is an asteroid \citep{HuntFoxWil1986, Babadzhanov1994, Bottke2002, Licandro2007}, a comet \citep{Fox1982, Ryabova2007, Ryabova2021}, or a member of a new intermediate class of objects \citep{JewittHsieh2022}. One of the most puzzling characteristics of Phaethon is its small and rarely observed activity, which is deemed insufficient to explain the total mass of the Geminides. In general, the nature of the mass emanation from Phaethon is not fully understood, despite observations running over the past four decades. 

The eccentric orbit of Phaethon likely contributes to its activity. The asteroid heats up at perihelion (0.14 AU) and cools down away from it, which cracks its surface, generating new layers of regolith; that is, small grains that are blown away by radiation from the Sun or electrostatic forces \citep{JewittLi2010, Jewitt2012, Delbo2014, Kimura2022}. In addition, \cite{Kasuga2009, Masiero2021, Zhang2023} presented evidence of sodium-driven activity, while \cite{Granvik2016, MacLennanGranvik2024} related the small heliocentric distance with thermal disruptions as activity drivers.  \cite{Wiegert2020} noted that the thermal disintegration of low perihelion asteroids forms many meteoroid-sized objects, opening up the possibility that impacts with larger bodies cause their activity. For a more detailed explanation of the potential drivers of the activity of asteroids, we refer the reader to \cite{Jewitt2019} and \cite{Neslusan2015}.

\citet{Ohtsuka2006} proposed that Phaethon or some other larger body could have fragmented in the past, forming the so-called Phaethon-Geminides complex (PGC). \citet{JewittHsieh2006, Ohtsuka2006, Ohtsuka2008} identified two nearby $\sim 1$ km asteroids, 155140 (2005 UD) and 225416 (1999 YC), as possible PGC fragments. Among the three, Phaethon and 2005 UD (hereafter denoted UD) exhibit greater orbital and physical similarities \citep{JewittHsieh2006, Kinoshita2007, Devogele2020, MacLennan2021, Ishiguro2022}, whereas the connection to 1999 YC (hereafter YC) appears somewhat less convincing; it is slightly more separated from Phaethon and UD, and is redder in spectral color \citep{KasugaJewit2008}. \cite{Hanus2018} showed that the cluster may have formed during the rotational fission of a critically spinning parent body \citep{Scheeres2007, Pravec2010}, but did not exclude the possibility of a collisional or thermally driven origin of the 3200–155140–225416 triplet.

However, not all authors agree with the scenario where the asteroids have a common parent body. For example, \cite{Schunova2012, Schunova2014} searched for "genetically" related asteroids amongst near-Earth objects (NEOs), reporting that the orbital similarity between the PGC members is statistically insignificant. \cite{Ryabova2019} ran backward integration, and showed that the three orbits were too far apart in the past 5000 years, concluding that neither of the three asteroids belongs to PGC, while \cite{Kareta2021} reported a difference between the near-infrared reflectance spectra of Phaethon and UD, claiming their similarities are coincidental. 

Blue asteroids are uncommon among NEOs. As noted in \cite{Hanus2016}, it appears unlikely that two B-type asteroids on similar (highly inclined and eccentric) orbits could simply be the results of random coincidence.

In this work, we study a scenario where all three asteroids, Phaethon, UD, and YC, arrived at their current locations from the main belt via mean motion resonances (MMRs). The considered source destinations are the outer-belt asteroid family (2) Pallas, and two inner-belt families, (329) Svea and (142) Polana. The first choice is supported by the results of \cite{Leon2010} and \cite{Todo2018, Todo2017}, who observed a Pallas–-Phaethon dynamical link, and by the work of \cite{Licandro2007}, \cite{Leon2010}, \cite{Hanus2018}, \cite{Marsset2020}, \cite{Devogele2018}, who found similarities between Phaethon and (2) Pallas. The inner-belt PGC source in the (329) Svea and/or (142) Polana families is proposed by \citet{MacLennan2021} and \cite{Bottke2002}.

Our secondary goal is to investigate some contradictions in the literature regarding the dynamical origin of Phaethon. We ran calculations using different parameter settings, data sets, initial epochs, and software packages (Orbit9 and Rebound) and examined how these changes affect the existence and efficiency of the possible dynamical routes.

\section{Methodology}
\label{section2}

Our methodology includes the following steps:

- First, we identify possible source regions for the three considered asteroids (Phaeton, UD, and YC). As mentioned in Sect. 1, those regions are the outer-main-belt asteroid family (2) Pallas, and two inner-belt families (329) Svea and (142) Polana.
    
- Mean motion resonances are often considered to be the main drivers responsible for dynamical transport. Thus, we select the strongest MMRs in the mentioned families and study their ability to deliver bodies to the targeted Apollo asteroids. In the (2) Pallas family, these are the 5:2 and 8:3 MMRs with Jupiter located at $a \sim 2.82$ au and $a \sim 2.70$ au, respectively. For (329) Svea and (142) Polana families, we select the 3:1 MMR with Jupiter at 2.5 au (the same resonance goes through both families at different inclinations).
    
- The next step is to identify the chaotic regions in the selected MMRs. We do this by mapping the resonances in the orbital plane of the largest (parent) body in the family, since our primary assumption is that particles are injected into the resonance from the vicinity of the parent body. 

- We calculate these maps using the MEGNO (Mean Exponential Growth of Nearby Orbits; \cite{Cincotta2003}) and FLI (Fast Lyapunov indicators; \cite{Froes1997a, Froes1997b, FGL2000}) chaos indicators. For MEGNO maps, we used the \textit{Python} Rebound\footnote{\href{https://rebound.readthedocs.io/en/latest/}{https://rebound.readthedocs.io/en/latest/}} module (\cite{Rein2012}), relying on the symplectic integrator WHFast (\cite{Wisdom1991}; \cite{Wisdom1996}; \cite{Rein2015}; \cite{Rein2019a}), while for FLI maps we used the Orbit9 software\footnote{\href{http://adams.dm.unipi.it/orbfit/}{http://adams.dm.unipi.it/orbfit/}}, which relies on the symplectic single-step method (implicit Runge–Kutta–Gauss) as a starter, and a multistep predictor for most of the propagation.
    
- We place a dense grid of initial N x N test objects in a given semi-major axis and eccentricity domain and calculate their corresponding MEGNOs (or/and FLIs). Test objects with higher MEGNOs are chaotic, while the stable ones have lower MEGNOs. The maps are produced by coloring each object according to its MEGNO value. In that way, the exact shape, location, and ---most importantly--- chaotic regions of the MMRs are drawn on the map.

- The best candidates for diffusion are particles that are located in the chaotic parts of the resonance. Selecting such objects directly from the map enables quick interaction with the resonance, which saves a significant amount of computational resources. 

- The next step is to track the orbital evolution of the selected test particles (TPs) for a long time. Here we used the Orbit9 and Mercurius integrators\footnote{Mercurius is a hybrid integrator very similar to the hybrid integrator in John Chambers' Mercury code (\cite{Chambers1999}). It uses a symplectic integrator and switches over to a high-order non-symplectic integrator during close encounters. Specifically, Mercurius uses the efficient WHFast and IAS15 (Integrator with Adaptive Step-size control, 15th order) integrators internally.} (\cite{ReinSpi2015}; \cite{Rein2019b}). 
    
- Finally, we count all objects that enter into some given neighborhood of the targeted asteroids during the integration time. The neighborhood encompasses only three orbital elements: semi-major axis, eccentricity, and inclination. In other words, we count all objects whose orbital elements (a, e, i) satisfy the condition:
    
\begin{equation}
|a - a_{t} | < \Delta a, |e - e_{t}| < \Delta e,  |i - i_{t} | < \Delta i,
\label{condition}
\end{equation}
    
where $a_{t}, e_{t}$, and $i_{t}$ are the semi-major axis, eccentricity, and inclination of the target asteroid, and $\Delta a, \Delta e,$ and $\Delta i$ are some small arbitrary values defining the neighborhood of the considered orbital elements.

\section{Results}

In this section, we first show the 5000 year dynamical maps obtained with  MEGNO and/or FLI for each studied resonance (5:2J, 8:3J, and 3:1J) within the asteroid families that we consider. We calculate their transportation ability to the asteroids Phaethon, UD, and YC over 5 Myr of orbital evolution based on 1000 test asteroids selected from their chaotic borders. We count the TPs that fulfill the condition (\ref{condition}) for each targeted asteroid, where $\Delta a = 0.1  au, \Delta e = 0.1$, and $\Delta i = 5^{\circ}$.

The calculations are performed using the osculating orbital elements and the ephemerides are taken from the JPL Horizons system\footnote{\href{https://ssd.jpl.nasa.gov/horizons.cgi}{https://ssd.jpl.nasa.gov/horizons.cgi}}. Our model includes all eight planets without their natural satellites and does not include the YORP or Yarkovsky nongravitational effects.

\subsection{The Pallas family}

\cite{Leon2010} were the first to notice a possible dynamic connection between the (2) Pallas family and the asteroid (3200) Phaethon. These authors observed a link acting via the 5:2 and 8:3 MMRs with Jupiter, and with a probability of 2\% in 100 Myr. Using the methodology described above and the Orbit9 integrator, \cite{Todo2018} observed a more convincing Pallas--Phaeton connection through both resonances, with a probability of close to 50\% in 5 Myr.
 
\cite{MacLennan2021} tracked the backward time evolution of the nominal Phaethon and UD orbits and their clones with the SWIFT RMVS integrator \citep{Levison1994}, and ruled out a dynamical connection to the Pallas family. Also \cite{Kovacova2022} did not register any particles from the 8:3 MMR on an orbit close to that of Phaethon. \cite{Kovacova2022} used the Rebound software package and a methodology similar to \cite{Todo2018}, but the test asteroids were placed in the orbital plane of the lower inclined asteroid (1) Ceres.

The (2) Pallas family is intertwined with many other resonances: for example, the 13:5J and 7J:2S:3A at $a \sim 2.705$ au, 1J:3S:1A at $a \sim 2.752$ au, and even 1:1 MMR with Ceres at 2.767 au. Several secular resonances (SRs) are present as well (see \cite{Carruba2010, TodoNov2015, Gallardo2016} for a more detailed study).
In this section, we focus on the strongest ones, the 5:2J and 8:3J MMRs. In addition to studying their transport ability toward Phaeton, UD, and YC, we also investigate the difference between results obtained with Orbit9 and Rebound/Mercurius, and between results obtained with different sets of TPs, integration steps (for the 5:2J MMR) and initial epochs (for the 8:3J MMR).

\begin{figure}[h!]
    \includegraphics[width=\columnwidth, height=0.85\columnwidth]{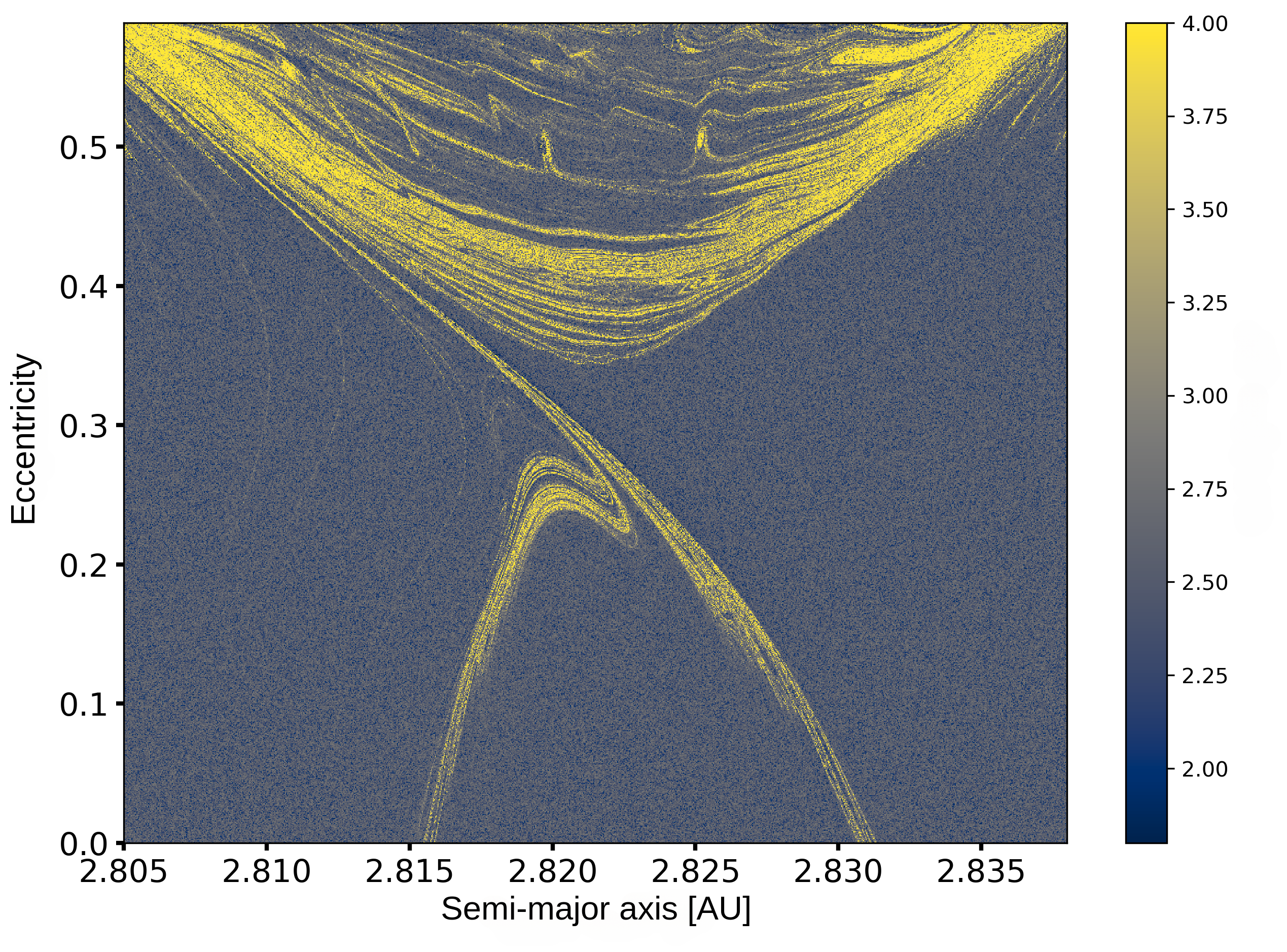}
    \caption{MEGNO map of the 5:2J MMR. It is calculated for 5000 years in the orbital plane of (2) Pallas for the epoch of 30 September 2012. Chaotic particles with larger MEGNOs are yellow, while relatively stable parts with lower MEGNOs are blue. As the asteroid (2) Pallas is located at $(a, e) = (2.77, 0.23)$, and the members of the family have $e \in [0.19, 0.364]$, particles for diffusion are selected along the yellow structures in the eccentricity range of the family from the narrowest part of the resonance. }
    \label{Fig1}
\end{figure}

\textit{The 5:2 MMR with Jupiter.} This resonance is located in the outer border of the Pallas family at 2.823 au. The MEGNO map of this resonance is shown in Fig. \ref{Fig1}, where a grid of $1000 \times 1000$ initial test objects is placed in the $[a, e] = [2.805, 2.838] \times [0, 0.589]$ domain, while the four remaining orbital elements, namely inclination $i$, the longitude of the node $\Omega$, the longitude of the pericenter $\omega$, and the mean anomaly $M$, are equal to the corresponding elements of Pallas for the epoch of 30 September 2012 that is, $\{i, \Omega, \omega, M\} = \{34.8^{\circ}, 173.09^{\circ}, 310^{\circ}, 248.97^{\circ}\}$. To compare FLI and MEGNO maps, we use the same epoch as in \cite{Todo2018}. For each particle, we calculate the MEGNO value for 5000 years and color it according to its stability properties: chaotic particles are yellow, and stable ones are blue. 

Generating a map of one million test objects and integrating over 5000 years with an integration time step of $\Delta t = 1$ year takes approximately 2.5 hours on a Ubuntu 22.04 LTS, AMD Ryzen 9, with 128 GB RAM. Running the FLI map with Orbit9 for the same configuration and parameters, takes significantly more time, at least several days. 

The observed chaotic structures in Fig. \ref{Fig1} match the structures in the FLI map from Todorovi\'c \citeyearpar[see Fig. 1 in the paper]{Todo2018}, although with somewhat less clarity and without the numerous weak MMRs outside the 5:2J resonance. The structures inside the 5:2J MMR are identical and are not even shifted along any of the axes; we regard this as a sort of validation of the mapping results.

Objects for diffusion are selected from the chaotic edges of the resonance within the eccentricity range of the Pallas family, that is, for $e\in [0.19, 0.364]$. In Fig. \ref{Fig1}, this corresponds to the narrowest part of the structure.  Whether Pallas ---as one of the most massive asteroids--- influences this unusual hourglass shape of the resonance remains an open question. Nevertheless, 1000 TPs were selected from the narrowest part and further tracked for 5 Myr. 
 
\begin{figure*}
\centering
\begin{minipage}[t]{\textwidth}
 \textbf{(a)}
\end{minipage}
\begin{minipage}[t]{0.45\textwidth}
  \includegraphics[width= \linewidth, height= 0.7\linewidth]{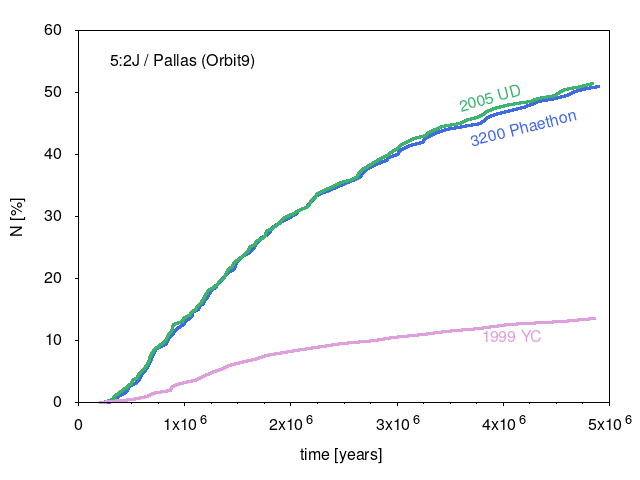}
\end{minipage}%
\begin{minipage}[t]{0.45\textwidth}
  \includegraphics[width=\linewidth,height= 0.7\linewidth]{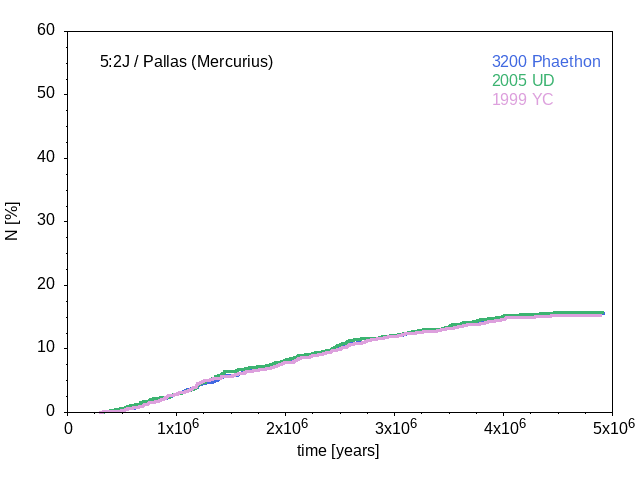}
\end{minipage}
\vskip -0.70cm
\begin{minipage}[t]{\textwidth}
\textbf{(b)}
\end{minipage}
\begin{minipage}[t]{0.45\textwidth}
  \centering
  \includegraphics[width=\linewidth, height= 0.7\linewidth]{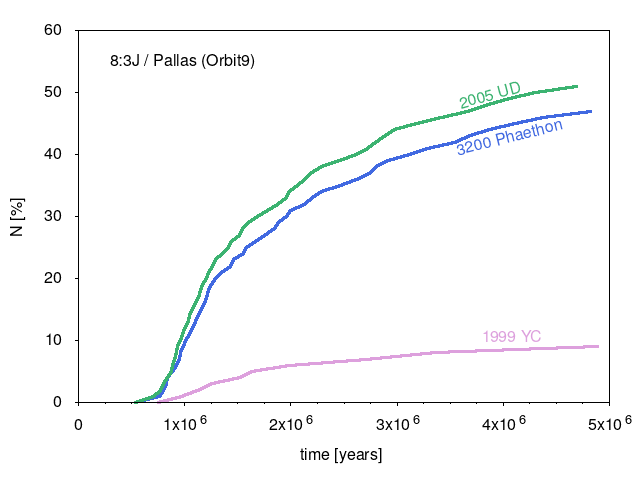}
\end{minipage}%
\begin{minipage}[t]{0.45\textwidth}
  \centering
  \includegraphics[width=\linewidth, height= 0.7\linewidth]{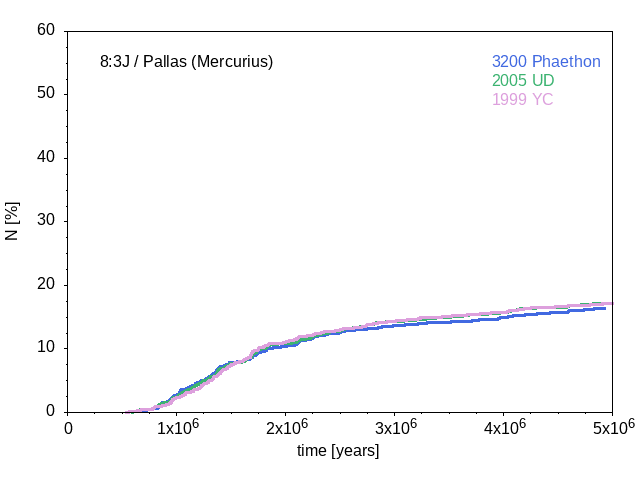}
\end{minipage}
\vskip -0.70cm
\begin{minipage}[t]{\textwidth}
\textbf{(c)}
\end{minipage}
\begin{minipage}[t]{0.45\textwidth}
  \centering
  \includegraphics[width=\linewidth,height= 0.7\linewidth]{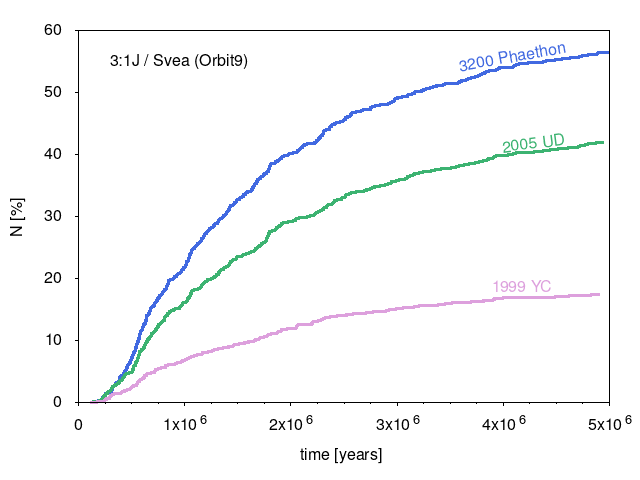}
\end{minipage}%
\begin{minipage}[t]{0.45\textwidth}
  \centering
  \includegraphics[width=\linewidth, height= 0.7\linewidth]{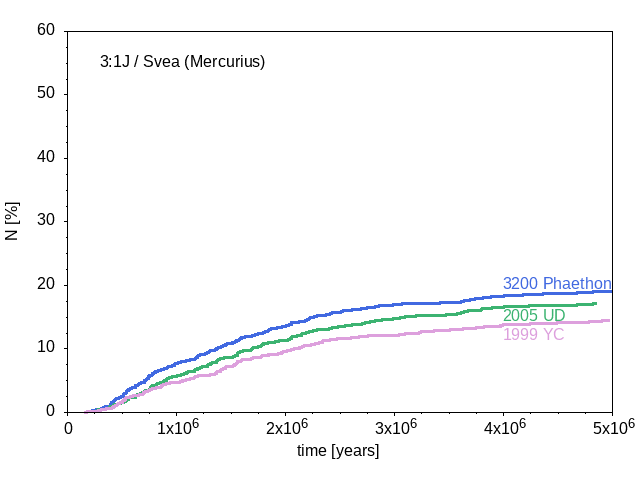}
\end{minipage}
\vskip -0.70 cm
\begin{minipage}[t]{\textwidth}
\textbf{(d)}
\end{minipage}
\begin{minipage}[t]{0.45\textwidth}
  \centering
  \includegraphics[width=\linewidth, height= 0.7\linewidth]{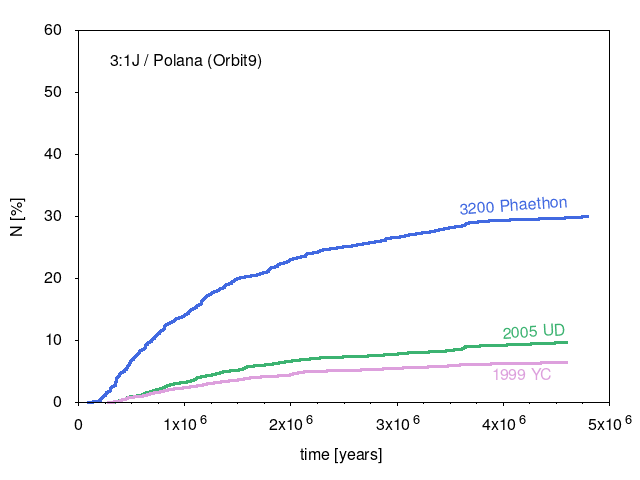}
\end{minipage}%
\begin{minipage}[t]{0.45\textwidth}
  \centering
  \includegraphics[width=\linewidth, height= 0.7\linewidth]{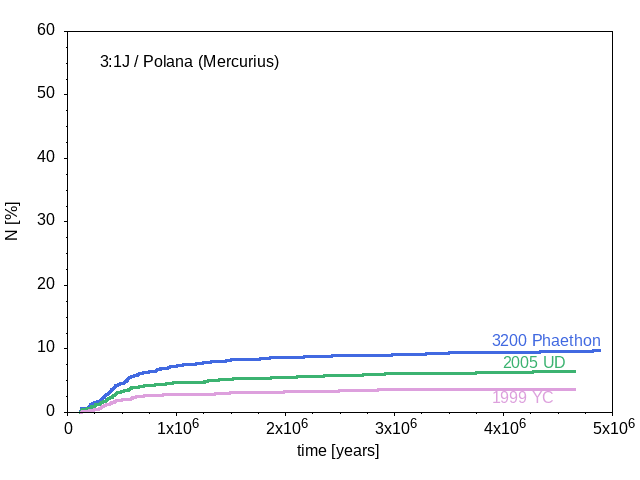}
\end{minipage}
\vskip -0.35 cm
\caption{The 5 Myr migration rate to the asteroids Phaethon (green), UD (blue), and YC (pink). The upper four panels show the transportation from the asteroid family (2) Pallas via the 5:2J (a) and 8:3J (b) MMRs. Panels (c) and (d) present migrations from the (329) Svea and (142) Polana families via the 3:1J resonance. On the left are the results obtained with Orbit9, and on the right those with Rebound/Mercurius. Orbit9 displays a noticeably greater transfer efficiency, particularly for asteroids Phaethon and UD, with around 50\% of TPs arriving from the resonances in Pallas and Svea families. In Mercurius, the 5:2J and 8:3J MMRs show almost identical transportation to all three asteroids: the curves on the right in (a) and (b) are overlapping. These lines are separated in the lower two panels for Mercurius (panels on the right side of panels (c) and (d)), showing the highest arrival efficiency to Phaethon, a slightly lower efficiency to UD, and the lowest to YC. More details are given in the main text and Tables \ref{Table 1},\ref{Table 2}, and \ref{Table 3}.}
\label{Fig2}
\end{figure*}

For this data set, we ran the calculation four times. First, using Orbit9 with an adaptive time step, while the other three runs are performed with the Mercurius integrator for $\Delta t = \{1, 0.1, 0.01\}$ years, switching to a shorter integration step during the close encounter. The results of their arrivals at Phaethon, UD, and YC are presented in Table \ref{Table 1} and in the top two panels of Fig. \ref{Fig2}. 

Unlike in maps, there is a systematic difference between the results obtained with Orbit9 and those obtained with Mercurius regarding transportation. According to Orbit9, 51.5\% of TPs reached Phaeton, 51.9\% reached UD, while the connection to YC was the least efficient with, 13.6\% TPs reached this asteroid.

Mercurius showed a less convincing delivery, with about $\sim 15\%$ of TPs reaching each of the three asteroids while using the time step  $\Delta t = 1$. When decreasing the time step of integration, the percentages of arrival objects are also expected to decrease, which was confirmed only in the case of YC. Contrary to such expectations, if we decrease the time step to $\Delta t = 0.1$, the connection to Phaeton becomes more effective, with 20\% arrival. This had little impact on the number of asteroids that reached UD, which remained at approximately 15\%. The case with $\Delta t = 0.01$ saw the fewest TPs arriving at their destinations, with only 7.4\% reaching Phaeton; a negligible number (only 4) reaching UD, and none at all reaching YC. A certain number of particles visited all three asteroids, 9.8\% of them for  $\Delta t = 1$, 7.3\% for $\Delta t = 0.1$, and only 0.4\% for $\Delta t = 0.01$ (obtained with Mercurius and not shown in the table).

The first arrival times for $\Delta t = 1$ are about $t \sim 300$ Kyr for all three asteroids. When shortening the integration step, the transportation times increase, most notably for UD, where they almost triple, from 313.66 Kyr for $\Delta t = 1$ to 988.58 Kyr for $\Delta t = 0.01$. Median arrival times do not show a significant difference when changing $\Delta t$, being about $t \sim 2$ Myr for all cases. Again, Orbit9 gives different results, somewhat faster transport with first arrivals of about $t \sim 200$ Kyr and $t \sim 1.6$ Myr for median times.

\begin{table*}[h!]
\vskip 0.3cm
\caption{Results of a 5 Myr transport from the 5:2J resonance for Phaethon, UD, and YC.} 
\centering { 
\small
\begin{tabular}{l l c c c c c c }
 & & & & & & &  \\
 & & {(3200) Phaethon} & & {(155140) 2005 UD} & & {(225416) 1999 YC} &  \\
 \arrayrulecolor{gray}\hline
 \rowcolor{blue!20} \% of arrived objects &  &  &  &  &  &  &   \\ 
\hline
\multirow{3}{*}{\rotatebox[origin=c]{90}{Orbit9}}   
  &  &  &  &  &  &  &  \\  
  & Adaptive time-step & 51.50 \% &  & 51.90\% &  & 13.60\%  &  \\ 
  &  &  &  &  &  &  &   \\                    
\hline
  &  &  &  &  &  &  &   \\  
\multirow{3}{*}{\rotatebox[origin=c]{90}{Rebound}} 
  & {$\Delta t = 1$} & 15.70\% & & 15.90\% & & 15.40\% & \\  
  & {$\Delta t = 0.1$} & 20.00\% &  & 15.50\% &  & 8.40\% &  \\
  & {$\Delta t = 0.01$}  & 7.40\% &  & 0.04\% &  & - &  \\
  &  &  &  &  &  &  &   \\
\hline
\rowcolor{blue!20} $t_{first\ arrival} [Kyr]$ &  &  &  &  &  &  &   \\ 
\hline
\multirow{3}{*}{\rotatebox[origin=c]{90}{Orbit9}} &  &  &  &  &  &  &  \\ 
 & Adaptive time-step & 278.50 &  & 232.50 &  &  205.00 &   \\ 
 &  &  &  &  &  &  &   \\ 
\hline
 &  &  &  &  &  &  &   \\
\multirow{3}{*}{\rotatebox[origin=c]{90}{Rebound}} 
 &  $\Delta t = 1$ &  314.37 &  &  313.66 &  & 304.67 &   \\  
 &  $\Delta t = 0.1$  &  437.49 &  &  438.27 &  & 628.28 &  \\  
 &  $\Delta t = 0.01$ &  583.70 &  &  988.58 &  & - &   \\                          
 &  &  &  &  &  &  &   \\                                      
\hline 
\rowcolor{blue!20} $t_{median} [Myr]$ &  &  &  &  &  &  &   \\  
\hline
\multirow{3}{*}{\rotatebox[origin=c]{90}{Orbit9}} &  &  &  &  &  &  &  \\ 
 & Adaptive time-step & 1.674 &  & 1.687 &  & 1.566 &   \\ 
 &  &  &  &  &  &  &   \\
\hline
 &  &  &  &  &  &  &   \\  
\multirow{3}{*}{\rotatebox[origin=c]{90}{Rebound}} 
 & $\Delta t = 1$ & 1.973 &  & 1.949 &  & 1.967 &   \\  
 & $\Delta t = 0.1$ & 2.047 &  & 2.162 &  & 2.163 &   \\  
 & $\Delta t = 0.01$ & 2.182 &  & 2.108 &  & - &   \\  
 &  &  &  &  &  &  &   \\                   
\hline
\end{tabular}
\vskip 0.3cm
\begin{minipage}{\linewidth}
\textbf{Notes.} The initial conditions are taken from the chaotic edges of the resonance in Fig. \ref{Fig1}. The results are obtained with Orbit9 and Rebound/Mercurius for three different integration time steps: $\Delta t = \{1, 0.1, 0.01\}$ years. The upper rows show the percentage of objects entering a given neighborhood of Phaethon, UD, and YC. The middle rows show their first arrival times, and the bottom their median arrival times. These results are discussed in the main text.
\end{minipage}
}
\label{Table 1}
\end{table*}

The graphical representation of the migration rate via the 5:2 MMR is given in the top two panels of Fig. \ref{Fig2}, where the arrivals to Phaeton are colored blue, those to UD are green, and those to YC are pink. The right hand panel shows the results obtained with Mercurius for a time step $\Delta t = 1$, while the left panel shows the results obtained with Orbit9. In the Mercurius case, the flow of TPs to Phaeton, UD, and YC almost overlaps, with similar delivery to all three asteroids. In Orbit9, TPs towards UD and Phaeton are almost identical in number and delivery time. The number of TPs reaching YC is significantly lower, but this is the only aspect of the results where Mercurius and Orbit9 agree.

As an additional check, we took another set of 1000 objects from the chaotic border of the resonance and repeated the whole procedure. Almost identical transportation efficiency was observed for all three asteroids.

Calculations of orbits in Mercurius for an integration time step of $\Delta t = 1$ year take approximately 17 hours, while those for a time step of $\Delta t = 0.1$ years take about 40 hours, and those for a time step of $0.01$ years take 270 hours on a Ubuntu 22.04 LTS, AMD Ryzen 9, with 128 GB RAM. We tried to calculate orbits for even shorter time steps $\Delta t$, but since these calculations require over a month to finish, they were not performed. 

\begin{figure}[h]
    \includegraphics[width=\columnwidth, height=0.85\columnwidth]{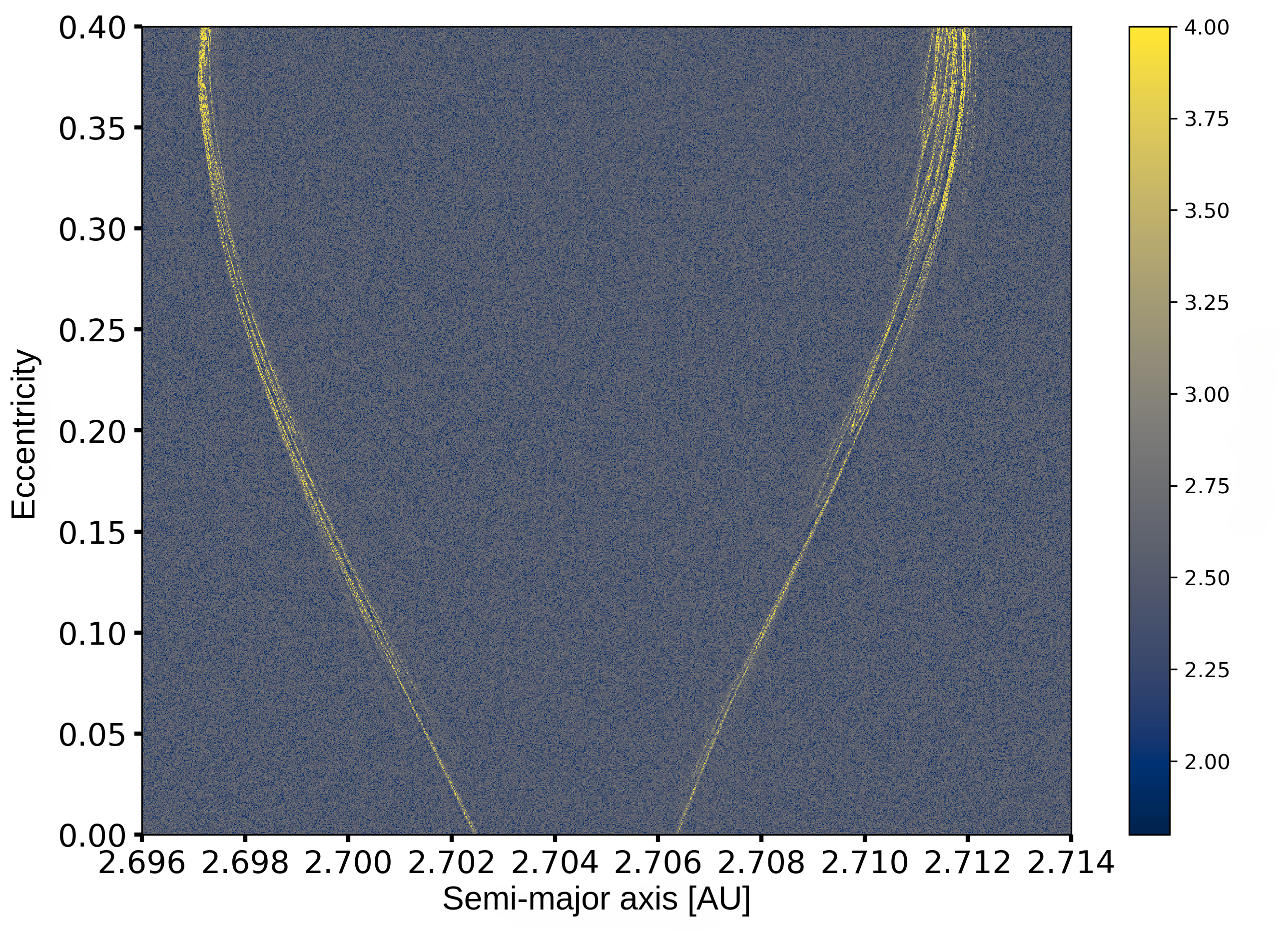}
    \caption{MEGNO map of the 8:3J MMR in the orbital plane of Pallas. Parameters are the same as in the Fig. \ref{Fig1}.}
    \label{Fig3}
\end{figure}

\textit{The 8:3 MMR with Jupiter.} This is another strong driver from the Pallas family, and is located on its inner edge at approximately $\sim 2.7$ au. The MEGNO map of the resonance is given in Fig. \ref{Fig3}; it covers the area $[a, e] = [2.696, 2.714] \times [0, 0.4]$, and is calculated for the same epoch and parameters as in Fig. \ref{Fig1}. The structures on the map closely agree with those on the FLI map obtained with Orbit9 (see Fig. 3 in \cite{Todo2018}). We also produced a MEGNO map for the epoch of 26 August 2022; it is not shown here because it is almost identical to the one in Fig. \ref{Fig3} (it is slightly shifted to the left for $a \sim 0.002$ au).
For both maps, we take 1000 TPs from the chaotic edge of the resonance within the family eccentricity range $e\in [0.19, 0.364]$, track them for 5 Myr, and check their arrivals to Phaeton, UD, and YC. The results are presented in Table \ref{Table 2}, where again, for Orbit9 they are obtained using an adaptive time-step, while for Mercurius we used an integration time step of $\Delta t = 1$ year.

For the epoch of 30 September 2012, Orbit9 shows that 47.5\% of TPs reach Phaeton, 52\% reach UD, and only 9.1\% reach YC. Mercurius shows a lower efficiency for the same epoch, with around $\sim 17\%$  approaching Phaethon, UD, and YC. For 26 August 2022, this latter suggests a $\sim 5 - 6\%$ higher transport efficiency to the targeted asteroids. Also, for this resonance, the same objects were seen to approach all three asteroids, 11\% of them for the first epoch and 16.2\% for the second (these data are not shown in the table).

In regards to the first arrival times, Orbit9 shows values of $\sim 530$ Kyr for both Phaethon and UD and a somewhat longer first delivery to YC at $757.5$ Kyr. For the same epoch of 30 September 2012, Mercurius gives almost identical first arrivals for Phaethon and UD, of namely $\sim 599$ Kyr, but a slightly shorter time for YC of $\sim 542$ Kyr. 

For 26 August 2022, YC and UD have identical first arrivals of $\sim 481$ Kyr, and a slightly longer time to Phaethon of $553.44$ Kyr. On average, the deliveries from the 8:3J MMR are around 0.5 Myr for all three asteroids in both software packages, although these times are somewhat shorter for the second epoch. 

For 30 September 2012, we find median delivery times with Orbit9 of $t \sim 1.5$ Myr for the three targets (the delivery to UD is slightly faster with $t \sim 1.4$ Myr), and in the case of Mercurius, median delivery times are slightly above $1.6$ Myr in all three cases. For 26 August 2022, we find median times that are on average $300$ Kyr shorter than for the first epoch (in Mercurius), of namely $t \sim 1.3$ Myr. 

A graphical representation of the deliveries via the 8:3J MMR is given in Fig. \ref{Fig2}b. Integration in Orbit9 (left) shows more efficient deliveries than Mercurius (right), with most TPs reaching UD. Regarding the 5:2J resonance, Mercurius gives almost identical results for all three asteroids. On average, the first arrival times from the 8:3J MMR are nearly two times longer than for the 5:2J MMR, while the median times appear shorter.

Using a similar method, \cite{Kovacova2022} did not confirm the connection between the 8:3J resonance and Phaeton. The discrepancy between our results and theirs most likely appears because these latter authors mapped the resonance in the orbital plane of (1) Ceres, at an inclination of 10 degrees, from which transport is likely to be less efficient. 

\begin{table*}[h!]
\vskip 0.3cm
\caption{Results of 5 Myr of transport via the 8:3J resonance for Phaethon, UD, and YC.} 
\centering { 
\small
\begin{tabular}{l l c c c c c c }    
 & & & & & & & \\
 & & {(3200) Phaethon} & & {(155140) 2005 UD} & & {(225416) 1999 YC} &  \\
 \arrayrulecolor{gray}\hline
 \rowcolor{blue!20} \% of arrived objects &  &  &  &  &  &  &   \\ 
\hline
\multirow{3}{*}{\rotatebox[origin=c]{90}{Orbit9}}   
  &  &  &  &  &  &  &   \\  
  & 30 September 2012 & 47.5\% &  & 52.0\% &  & 9.1\% &   \\ 
  &  &  &  &  &  &  &   \\                    
\hline
  &  &  &  &  &  &  &   \\  
\multirow{3}{*}{\rotatebox[origin=c]{90}{Rebound}}
  &  &  &  &  &  &  &   \\
  & 30 September 2012 & 16.50\% & & 17.30\% & & 17.30\% &   \\  
  & 26 August 2022 & 22.80\% &  & 23.80\% &  & 22.00\% &  \\
  &  &  &  &  &  &  &   \\
\hline
\rowcolor{blue!20} $t_{first\ arrival} [Kyr]$ &  &  &  &  &  &  &   \\
\hline
\multirow{3}{*}{\rotatebox[origin=c]{90}{Orbit9}} &  &  &  &  &  &  &   \\ 
 & 30 September 2012 & 534.0 &  & 532.5 &  &  757.5 &   \\ 
 &  &  &  &  &  &  &   \\ 
\hline
 &  &  &  &  &  &  &   \\ 
\multirow{3}{*}{\rotatebox[origin=c]{90}{Rebound}}
 &  &  &  &  &  &  &   \\
 &  30 September 2012 &  599.90 &  &  599.39 &  & 542.16 &   \\  
 &  26 August 2022 &  553.44 &  &  481.60 &  & 481.75 &   \\                           
 &  &  &  &  &  &  &  \\                                     
\hline 
\rowcolor{blue!20} $t_{median} [Myr]$ &  &  &  &  &  &  &   \\  
\hline
\multirow{3}{*}{\rotatebox[origin=c]{90}{Orbit9}} &  &  &  &  &  &  &   \\ 
 & 30 September 2012 & 1.532 &  & 1.443 &  & 1.56 &   \\ 
 &  &  &  &  &  &  &   \\
\hline
 &  &  &  &  &  &  &   \\ 
\multirow{3}{*}{\rotatebox[origin=c]{90}{Rebound}}
 &  &  &  &  &  &  &   \\
 & 30 September 2012 & 1.630 &  & 1.672 &  & 1.665 &   \\  
 & 26 August 2022 & 1.384 &  & 1.391 &  & 1.385 &  \\ 
 &  &  &  &  &  &  &   \\                   
\hline    
\end{tabular}
\vskip 0.3cm
\begin{minipage}{\linewidth}
\textbf{Notes.} The initial conditions are taken from the chaotic edges of the resonance from Fig. \ref{Fig3}. The table shows the percentage of arrived objects, and their first and median arrival times to the Phaethon, UD, and YC, for two different epochs, 30 September 2012 and 26 August 2022, and two different integrators, Orbit9 with an adaptive time-step, and Rebound with $\Delta t = 1$ year integration step. Asteroids taken in the second epoch are somewhat more efficiently transferred.
\end{minipage}
}
\label{Table 2}
\end{table*}

\subsection{Svea and Polana families}

A possible inner belt origin of Phaethon was first suggested by \cite{Bottke2002}. Later, \cite{MacLennan2021} confirmed this assumption for both Phaethon and UD and listed (329) Svea and (142) Polana families as their potential sources. The escape routes identified by the authors were the 3:1J MMR  and the secular $\nu_6$ resonance, which occurs when the asteroid and Saturn have equal precession frequencies. 

Below, we present a study of the ability of the 3:1J MMR to deliver asteroids to the three Apollo targets from the (329) Svea and (142) Polana families. Although the $\nu_6$ resonance is one of the most important resonances in the main belt and has a significant role in capturing \citep{Huaman2018, Carruba2022} and delivering objects from the Svea family into the region of the tree-targeted asteroids \citep{MacLennan2021}, we did not include it in our analysis. For the clarity of MEGNO/FLI maps, it is crucial to have short integration times, enough to register MMRs, their edges, and fine structures. By extending time, the sharpness will be lost and MMRs will be drawn into chaos. On the other hand, such short times are insufficient to detect any SRs, not even $\nu_6$, because a typical period of oscillations in SRs takes a few million years, several orders of magnitude longer than the times used to create these maps.

\begin{figure*}[h!]
\centering
\begin{minipage}{\textwidth}
\includegraphics[width= 0.5\linewidth]{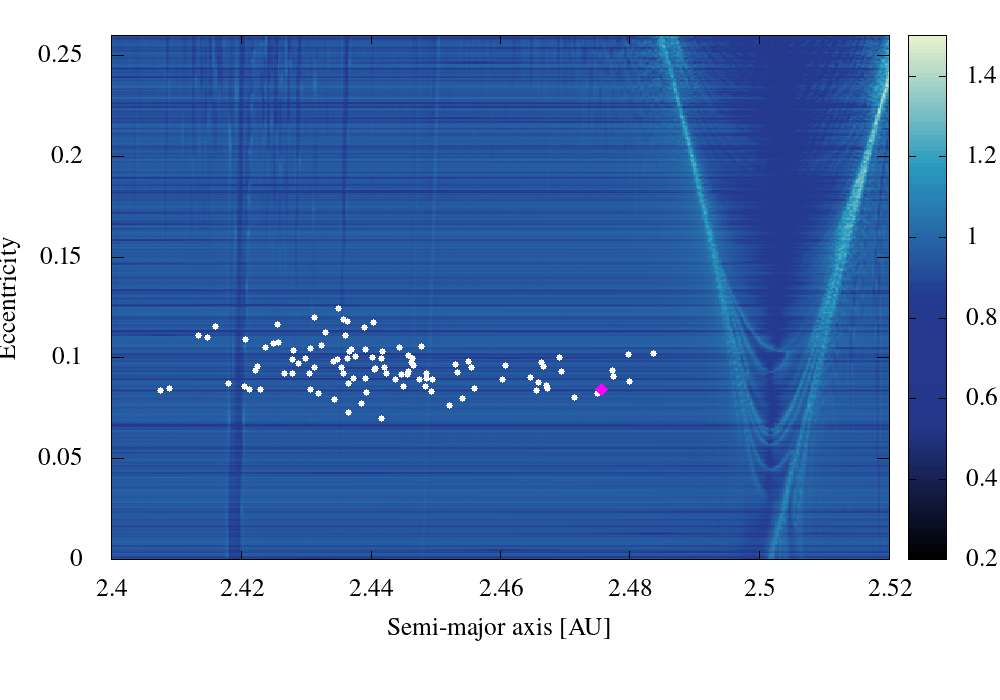}
\includegraphics[width= 0.5\linewidth]{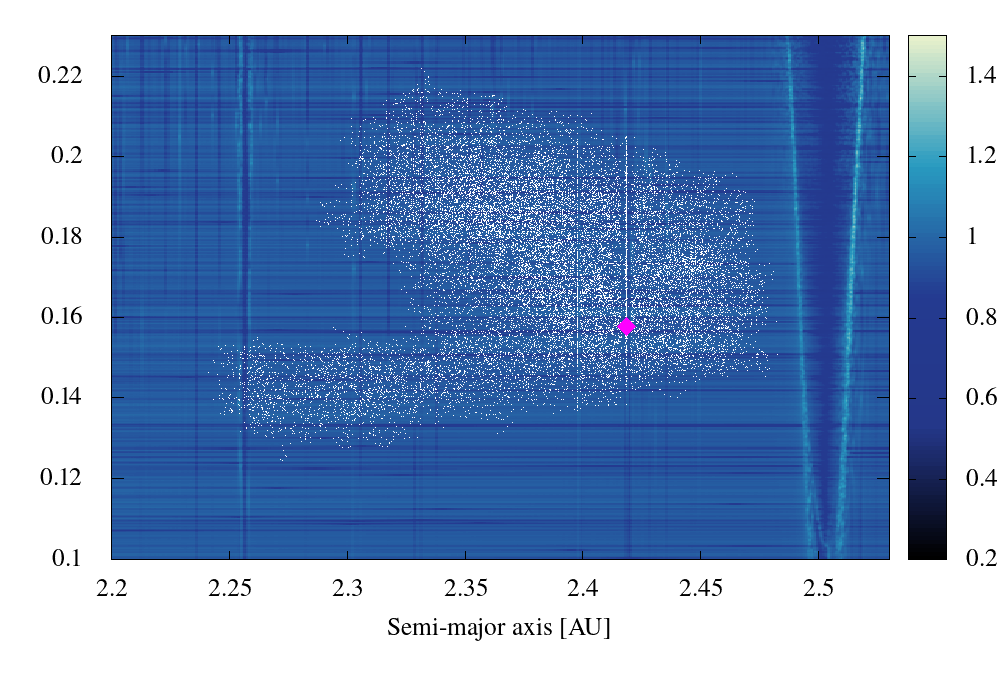}
\end{minipage}
\caption{FLI map of the 3:1J MMR in the orbital planes of (329) Svea (left) and (142) Polana (right). The $[a,e]$ domains are extended to the corresponding family ranges. The members of the families are plotted as white dots, while the position of the asteroids Svea (left) and Polana (right) appear as pink diamond-like shapes. In addition to the 3:1J MMR at 2.5 au, both charts show the presence of other weaker MMRs, among which the 1:2 MMR with Mars at $a=2.42$ au appears very successful in trapping asteroids from the Polana family, including the Polana asteroid itself. The migration results from the 3:1J MMR to the asteroids Phaethon, UD, and YC are given in the main text in Table \ref{Table 3} and panels (c) and (d) in Fig. \ref{Fig2}.}
\label{Fig4}
\end{figure*}

\textit{The 3:1 MMR with Jupiter.} It is located at about 2.5 au, on the outer edge of both (329) Svea and (142) Polana families at different inclinations: for Svea, this is at about $i \sim 16^{\circ}$, for Polana this is at  $i < 6^{\circ}$. Figure \ref{Fig4} shows two FLI maps of the 3:1J resonance calculated with Orbit9 at both inclinations and in the orbital planes of the two asteroids, where the $[a, e]$ range is extended to the domain of the two families. We used the osculating orbital elements for the epoch of 13 November 2019, and a $500 \times 500$ grid. Chaotic and stable parts are marked with lighter and darker shades of blue. MEGNO maps were also calculated, but we show only the FLI maps because they contain much more detail and are of greater clarity. 

On the left panel, we show the 3:1J MMR in the framework of the (329) Svea family for $[a, e] = [2.4, 2.52] \times [0, 0.26]$ with ($i$, $\Omega$, $\omega$, $M)=(15.83, 178.44, 50.12,  231.71)$. The 3:1J resonance is visible at about 2.5 au as a large V-shaped form. Other weak MMRs are also visible, mostly between $a=2.42$ and $a=2.44$  au; additional studies not shown here revealed that most of these are MMRs with Mars. As mentioned above, the $\nu_6$ resonance is not visible on the map, although its presence in the family affects almost 50\% of its members (see \cite{Huaman2018} for a more detailed study of the interaction between the Svea family and the $\nu_6$ resonance).

The positions of the family members are marked with white dots, while the parent body (329) Svea is marked with a pink diamond-like shape\footnote{Members of the (329) Svea family are generated by the \textit{Asteroid Families Portal} \citep{Radovic2017}, available at \href{http://asteroids.matf.bg.ac.rs/fam/index.php}{http://asteroids.matf.bg.ac.rs/fam/index.php}}. 

On the right, we show the 3:1J resonance in the domain of the (142) Polana family for $[a, e] = [2.2, 2.55] \times [0.1, 0.23]$ with the remaining elements equal to the corresponding elements of the asteroid 142 Polana; $(i, \Omega, \omega, M)=(2.24, 291.26, 290.06, 97.25)$. Members of the family ---more than 19000 of them--- taken from \cite{Nesvorny2020} and \cite{NesvornyBrCar2015}, are plotted as white dots, and the asteroid 142 Polana is shown as a pink diamond-like shape\footnote{For the (142) Polana family members we used \textit{Nesvorny HCM Asteroid Families} \citep{Nesvorny2020, NesvornyBrCar2015}, available at \href{https://sbn.psi.edu/pds/resource/nesvornyfam.html}{https://sbn.psi.edu/pds/resource/nesvornyfam.html}}. In addition to the 3:1J MMR, the map captures numerous weak resonances visible as thin vertical lines. In this paper, we do not examine the properties of the Polana family (see \cite{Cellino2001Icar, Walsh2013Icar} for a more detailed study), but we note that some of these weak MMRs appear to be very successful in capturing bodies. The largest concentration of captured family members (including the large asteroid 142 Polana) is at $a\sim 2.42$ au, in the 1:2 MMR with Mars. Additional checks with the \textit{Python} code in \cite{Smirnov2023AC} show that the asteroid 142 Polana librates in this resonance. Family members fit well into the dynamic background in the FLI map, even though they are given in synthetic proper elements and the map is presented in the osculating orbital elements.

Again, 1000 chaotic TPs are selected from the borders of the 3:1J MMR from both maps in Fig. \ref{Fig4}, within the eccentricity range of the families; for (329) Svea, this is within $e\in [0.06, 0.14]$, and for (142) Polana, this is within $e\in [0.14, 0.2]$.  The results of their approaches to Phaethon, UD, and YC during the 5 Myr integration time are shown in Table \ref{Table 3}. We ran integration in Orbit9 and Mercurius with a time-step of $\Delta t = 1$ year.

Table \ref{Table 3} shows that (329) Svea family members have a greater probability of reaching the three targets than those from the (142) Polana family. In Orbit9, these respective probabilities are 56.6\% compared to 30.1\% reaching Phaethon, 42.1\% compared to 9.7\% reaching UD, and 17.5\% compared to 6.6\% reaching YC. For Mercurius, this probability is 19.2\% compared to 9.8\% arriving at Phaethon, 17.2\% compared to 6.5\% arriving at UD, and 14.6\% compared to 3.8\% arriving at YC. There were particles taken from the 3:1J resonance that visited all three asteroids; this amounted to 10\% of TPs from the Svea family, and only 2.4\% from Polana (obtained with Mercurius and not shown in the Table \ref{Table 3}).

When looking at the first arrival times, all three asteroids show similar values if placed in the Svea family: arrival takes $120-140$ Kyr in Orbit9 and $160-180$ Kyr in Mercurius. The first arrival times from the Polana family range from 90 to 320 Kyr in the case of Orbit9 and are around 120 Kyr according to Mercurius.

Orbit9 gives similar median arrival times ranging within $1 - 1.3$ Myr for both families. In Mercurius, median times for Svea are $ 1.3 - 1.5$ Myr, and are about 2-3 times lower for Polana with $t \sim 0.5$ Myr. Polana has the lowest efficiency in terms of the number of transported objects, but the fastest deliveries. 

It should be mentioned that we also examined the transportation abilities of some of the weaker resonances in both families (using the Mercurius integrator), but the results were insignificant. For example, in the Polana family, the 7:2 MMR with Jupiter at $a \sim 2.253$ au gives a result of only 1 object out of 1000 (0.001\%) arriving in the vicinity of Phaethon and no objects arriving at UD or YC, while other weaker MMRs in both Svea and Polana families did not result in any objects approaching the targeted asteroids within 5 Myr. 

Resonances with Mars in the orbital plane of the Polana family seem to be very successful in capturing asteroids (as seen in the right panel of Fig.\ref{Fig4}) but not very efficient when it comes to delivering them, at least to the NEO region, as we see no objects approaching the asteroids there. However, in the Svea region, we find that the 1:2 MMR with Mars at $a\sim 2.42$ au delivered about 1.5\% test asteroids to Phaethon, 1.1 \% to UD, and 1.4 \% to YC. About 0.9\% of the TPs from this weak resonance reached all three asteroids. 

\begin{table*}[h!]
\vskip 0.3cm
\caption{Results of a 5 Myr integration on transport to Phaethon, UD, and YC.} 
\centering {
\small
\begin{tabular}{l l c c c c c c }
 & & & & & & &  \\
 & & {(3200) Phaethon} & & {(155140) 2005 UD} & & {(225416) 1999 YC} &  \\
 \arrayrulecolor{gray}\hline
 \rowcolor{blue!20} \% of arrived objects &  &  &  &  &  &  &   \\ 
\hline
\multirow{4}{*}{\rotatebox[origin=c]{90}{Orbit9}}   
  &  &  &  &  &  &  &   \\ 
  & (329) Svea     & 56.6\% &  & 42.1\% &  & 17.5\% &   \\ 
  & (142) Polana   & 30.1\% &  & 9.7\% &  & 6.6\% &  \\ 
  &  &  &  &  &  &  & \\                    
\hline
  &  &  &  &  &  &  &   \\  
\multirow{3}{*}{\rotatebox[origin=c]{90}{Rebound}}
  &  &  &  &  &  &  &  \\
  & (329) Svea   & 19.20\% & & 17.20\% & & 14.60\% &  \\
  & (142) Polana  & 9.80\% & & 6.50\% & & 3.80\% &  \\
  &  &  &  &  &  &  &   \\
\hline
\rowcolor{blue!20} $t_{first\ arrival} [Kyr]$ &  &  &  &  &  &  &  \\
\hline
\multirow{4}{*}{\rotatebox[origin=c]{90}{Orbit9}} 
 &  &  &  &  &  &  &   \\ 
 & (329) Svea   &  141.2 &  & 138.6 &  &  120.0 &  \\
 & (142) Polana & 90.5 &  & 322.0 &  &  282.3 &   \\ 
 &  &  &  &  &  &  &   \\ 
\hline
 &  &  &  &  &  &  &   \\ 
\multirow{3}{*}{\rotatebox[origin=c]{90}{Rebound}}
  &  &  &  &  &  &  &   \\
  & (329) Svea   & 176.62 & & 178.83 & & 161.58 &   \\
  & (142) Polana & 115.11 & & 123.11 & & 124.59 &   \\
  &  &  &  &  &  &  &   \\                                  
\hline 
\rowcolor{blue!20} $t_{median} [Myr]$ &  &  &  &  &  &  &   \\  
\hline
\multirow{4}{*}{\rotatebox[origin=c]{90}{Orbit9}} 
 &  &  &  &  &  &  &   \\
 & (329) Svea   & 1.253 &  & 1.323 &  & 1.319 &  \\
 & (142) Polana & 1.061 &  & 1.343 &  &  1.320 &  \\ 
 &  &  &  &  &  &  &  \\
\hline
 &  &  &  &  &  &  &   \\  
\multirow{3}{*}{\rotatebox[origin=c]{90}{Rebound}}
  &  &  &  &  &  &  &   \\
  & (329) Svea   & 1.296 & & 1.435 & & 1.490 & \\
  & (142) Polana & 0.534 & & 0.475 & & 0.445 &  \\
  &  &  &  &  &  &  &   \\                    
\hline    
\end{tabular}
\vskip 0.3cm
\begin{minipage}{\linewidth}
\textbf{Notes.} The initial conditions are taken from the chaotic edges of the 3:1J resonance from Fig. \ref{Fig4}. The table shows the percentage of arrived objects and their first and median arrival times to the three asteroids from the (329) Svea and (142) Polana families. The results are obtained with Orbit9 and Rebound/Mercurius. On average, the delivery of asteroids from the Polana orbital plane is somewhat faster but less effective regarding the number of transported bodies.
\end{minipage}
}
\label{Table 3}
\end{table*}

\section{Discussion and Conclusions}

Our results show that Phaethon, UD, and YC could have arrived from the main belt. These connections are primarily attributed to low-order MMRs with Jupiter: 5:2J and 8:3J in the (2) Pallas family and the 3:1J bordering the (329) Svea and (142) Polana families. There is also a small possibility that weaker resonances, such as the 1:2 MMR with Mars, contribute to these deliveries. 

The effectiveness of the observed transits is highly dependent on the selection of parameters, but mostly on the integrator. Orbit9 and Rebound (i.e., FLI and MEGNO) give almost identical results in the production of dynamical maps. The transportation times also appear similar.  However, in the long-term integrations, we find a systematic discrepancy, where Orbit9 gives much higher transport probabilities than Rebound/Mercurius, especially for Phaethon and UD. 

Before we explain the possible reasons for this discrepancy, let us first reiterate the general delivery mechanism acting via MMRs. In the $10^5-10^6$ years of dynamical evolution, the orbits of test asteroids placed in MMRs become more eccentric, gradually increasing to planet-crossing values, which opens the possibility of one or multiple planetary encounters. Encounters are crucial in the delivery because they throw asteroids out of the resonance onto radically different trajectories. Orbit9 shows excellent performance regarding orbit integration, but it is not designed to handle very unstable cases, which would require regularization for close approaches (which is mentioned in the description of the program). Mercurius, on the other hand, is designed to handle close encounters: it switches to a high-order integration scheme during the approach \citep{Rein2019b}. Therefore, we treat the Mercurius results as more reliable for our case. 

Comparing the Mercurius results for the  5:2J, 8:3J, and 3:1J MMRs for the integration step of $\Delta t=1$ year, we get similar transport possibilities for all three MMRs: about 15-20\% of the selected TPs reached Phaethon, UD, and YC. Moreover, a smaller number of about 10-16\% TPs (except for the 3:1 MMR in the Polana family) experienced approaches to all three PGC members during the 5 Myr evolution. We treat the asteroids as point masses, neglecting their physical and rotational properties, but the particles reaching all three targets might suggest that the parent body of Phaethon broke apart during its chaotic evolution, after reaching small perihelion values. The disintegration of the parent body and its chaotic diffusion could happen simultaneously. 

Sensitivity to initial conditions appears in almost every definition of chaos. In our experiments, different sets of initial conditions taken along the chaotic border of the same resonance did not affect the results. However, changing the integration step from $\Delta t = 1$ to $\Delta t =0.01$ largely diminished the percentage of the delivered material and almost doubled the transportation time (see Table \ref{Table 1}). An unusual result for which we find no explanation is that more objects approached Phaethon for $\Delta t =0.1$ than for  $\Delta t =1$ years.

The phenomenon of chaos is not reversible, that is backward integration of chaotic orbits will probably not provide reliable solutions regarding their dynamic history. This may explain why the backward integration by \cite{Ryabova2019} denied any relationships between the PGC members, and why that by \cite{MacLennan2021} revealed no link to the (2) Pallas family. 

We also acknowledge the importance of Yarkovsky and YORP effects, because they can shorten diffusion times and affect our results to some degree.

In summary, we observed similar transport probabilities for the major resonances, except for the lower inclined 3:1J MMR in the Polana family. Nevertheless, we conclude that there is a greater chance that the Pallas family is the source region of the 3200–155140–225416 triplet, as it has two powerful transport channels, the 8:3J and 5:3J MMRs, and numerous weaker resonances, as reported by \cite{Carruba2010}, \cite{TodoNov2015}, \cite{Gallardo2016}, and therefore has at least two times greater chance of deporting bodies to the vicinity of Phaethon. The DESTINY+ \citep{Ozaki2022, Arai2024} flyby mission to Phaethon will have the last say on this matter.

\begin{acknowledgements}
This research was supported by the Ministry of Science, Technological Development and Innovation of the Republic of Serbia (MSTDIRS) through contract no. 451-03-66/2024-03/200002 made with the Astronomical Observatory (Belgrade) and contract no. 451-03-66/2024-03/200104 made with the Faculty of Mathematics at the University of Belgrade. 
\end{acknowledgements}

\end{document}